\documentclass{ws-p9-75x6-50}
\def\4He{$^4$He}
\def\cm{cm$^{-1}$ }   
\newcommand{\ket}[1]{\ensuremath{|#1\rangle}}
\newcommand{\bra}[1]{\ensuremath{\langle #1|}}
\newcommand{\avg}[1]{\ensuremath{\langle#1\rangle}}
\newcommand{\RR}{\ensuremath{\mathcal R}}

\begin{document}

\title{Structure and spectroscopy of doped helium clusters using quantum Monte Carlo techniques}

\author{Alexandra Viel}

\address{Lehrstuhl f\"ur Theoretische Chemie\\Technische Universit\"at M\"unchen \\85747 Garching, Germany}

\author{K. Birgitta Whaley}

\address{Dept. of Chemistry and K.S. Pitzer Center for Theoretical Chemistry \\University of California \\Berkeley, CA 94720-1460, USA}


\maketitle

\abstracts{
We present a comparative study of the rotational characteristics of various molecule-doped $^4$He clusters using quantum Monte Carlo techniques.
The theoretical conclusions obtained from both zero and finite temperature Monte Carlo studies confirm the presence of two different dynamical regimes that correlate with the magnitude of the rotational constant of the molecule, i.e., fast or slow rotors.
For a slow rotor, the effective rotational constant for the molecule inside the helium droplet can be determined by a microscopic two-fluid model in which helium densities computed by path integral Monte Carlo are used as input, as well as by direct computation of excited energy levels.
For a faster rotor, the conditions for application of the two-fluid model for dynamical analysis are usually not fulfilled and the direct determination of excitation energies is then mandatory.  
Quantitative studies for three molecules are summarized, showing in each case excellent agreement with experimental results.  
}

\section{Introduction}
Droplets of \4He provide a unique quantum environment that constitutes an ultra-cold and gentle matrix for high resolution spectroscopy\cite{toennies98} and for investigating the nature and dynamic consequences of quantum solvation for a wide range of impurities doped into superfluids.\cite{toennies01}
One of the most unusual features deriving from those experiments is the apparent free rotation of small molecules embedded inside these bosonic clusters at temperatures $T \sim 0.4 $~K\cite{kwon00}.
Rotational spectra obtained for a series of molecules possessing gas phase rotational constant $B_0$ in the range $0.01 - 50$ \cm \cite{toennies98,kwon00}, appear to fall into two dynamical regimes.
The heavier molecules (slow rotors) show a reduction in rotation constant of $\sim 60 - 80\%$, while lighter molecules (fast rotors) show a much smaller reduction of $0 - 30\%$.
In this article, we focus on the behavior of the three molecules SF$_6$, OCS, and HCN. 
The first two belong to the first dynamical regime of slow rotors, showing a similar reduction of $B/B_0 \sim 37\%$ in both cases\cite{toennies98}.
The third molecule, HCN, is a much faster rotor and the experimentally observed reduction of rotational constant $B/B_0$ is only $19\%$\cite{conjusteau00}.
For each of these molecules, accurate pair interaction potentials with helium exist, rendering quantitative calculations meaningful and reliable tests of quantum theories for molecular rotation in helium droplets.

\section{Methodologies}\label{methodologies}
Our theoretical studies of molecule-doped helium clusters are based both on $T=0$ methods diffusion Monte Carlo (DMC) and the projector Monte Carlo extension of this for excited states (POITSE), and on the finite temperature path integral Monte Carlo (PIMC) approach with incorporation of full quantum exchange symmetry\cite{kwon00}.
With DMC based techniques we evaluate directly the ground and excited energy levels of the doped cluster, focusing in this study on rotational excited states.
With the PIMC calculations we are able to compute finite temperature helium densities. These can then used as input for a two-fluid dynamical model that combines notions of adiabatic following with hydrodynamic estimates of superfluid response to molecular rotation\cite{kwon00}.
We provide in this section a short summary of how the quantum Monte Carlo methods are implemented for study of systems composed of a single molecule treated as a rigid body, together with $N$ helium atoms.

\subsection{Diffusion Monte Carlo}
The diffusion Monte Carlo technique for solving the Schr\"odinger equation for many-body systems is based on the imaginary time ($\tau=it/\hbar$) Schr\"odinger equation
\begin{equation}
\label{Eq_diffusion}
\frac{\partial\Psi({\cal R})}{\partial \tau} =
\sum_j^{\cal N} D_j \nabla_j^2\Psi({\cal R}) -
\left[V({\cal R})-E_{ref}\right]\Psi({\cal R}),
\end{equation}
where ${\cal R}$ is a vector in the ${\cal N}$-dimensional space, $D_j=\hbar^2/2m_j$ if the $j^{th}$ degree of freedom corresponds to a translation, and  $D_j=B_j=\hbar^2/2I_j$ if it corresponds to a rotation.
In the above equation, $E_{ref}$ is a constant defining the zero of the absolute energy scale, and $V$ is the potential.         
This formulation implies the use of Cartesian coordinates for the atom-like particles and for the center of mass of the rigid body, and the use of rotational angles around the principal axes of the rigid body. 
This diffusion-like equation is solved using random walks and a short time approximation for the Green's function\cite{whaley98}.
The effects of statistical noise on the results can be drastically reduced by introducing a guiding (or ``importance sampling'') function $\Psi_T$ that approximates the true solution $\Psi$ of Eq.(\ref{Eq_diffusion}), and by then rewriting the diffusion-like equation for the product function $f({\cal R}) = \Psi({\cal R})\Psi_T({\cal R})$ :
\begin{equation}
\label{Eq_diff_f}
\frac{\partial f({\cal R})}{\partial \tau} =
\sum_j^{\cal N} \left\{D_j \nabla_j^2f({\cal R})
-D_j \nabla_j\left[f({\cal R})F_j({\cal R})\right] \right\}
-\left[E_l({\cal R})-E_{ref}\right] f({\cal R}).
\end{equation}
Here $E_l({\cal R})=\Psi_T({\cal R})^{-1}\hat H \Psi_T({\cal R})$ is the local energy, and $F_j({\cal R})=\nabla_j \ln |\Psi_T({\cal R})|^2$ is the quantum force that controls the drift terms (second set of terms on right-hand side of Eq.~(\ref{Eq_diff_f})). 
We have recently shown~\cite{viel01} that one can use this importance sampling scheme for {\em all} degrees of freedom, including the rotation of the molecule when treated as a rigid body.
The average of the potential (in unbiased walks) or of the local energy (in biased, or ``importance sampled'' walks) over the random walk yields the ground state energy of the system. 
Whenever a short-time expansion for the Green's function is used, it is necessary to check that any time step bias is removed, or is at least smaller than the statistical noise.

With the incorporation of importance sampling into DMC it becomes possible to compute the helium density, or at least a mixed density $\langle \Psi_T |\rho |\Psi \rangle$, by simply binning the position of the helium atoms in a frame attached to the molecule during the random walk.
It is also possible to analyze the extent of {\em adiabatic following} of the molecular motion by the helium density, by artificially suppressing the rotational kinetic term of the dopant molecule in the Hamiltonian operator\cite{lee99,kwon00,patel01}.
Complete adiabatic following implies that the helium density in a frame attached to the molecule must be identical in both cases, i.e., with and without the incorporation of molecular rotation.
We have also proposed\cite{kwon00,patel01} a straightforward method to {\em quantify} the extent of the adiabatic following.
This can be done by evaluating the ratio of the densities along privileged axes in the molecule-helium system, with and without rotation:
 $Q(r) \equiv \left(\left[ \rho_{\rm saddle} / \rho_{\rm min} \right]_{\rm no \ rot}\right)/\left(\left[ \rho_{\rm saddle} / \rho_{\rm min}\right]_{\rm rot}\right)$.
When adiabatic following is significant,  $Q(r)$ is close to one.
In contrast, a value of $Q(r)$ close to zero indicates weak or minimal adiabatic following.

Just as in analysis of experimental spectra\cite{toennies98}, the rotational constants can be extracted by fitting the energy differences between different rotational states.
For spherical and linear rotors, the first energy difference between ground and first excited states is equal to $2B$.
This approach requires the explicit computation of excited state energies. 
Such computations can be done in an approximate way by making use of the widely employed fixed node approximation.
In this approximation, one imposes a predefined nodal surface on the excited state wave function by use of a trial wave function bias, 
Eq.~(\ref{Eq_diff_f}).
However, unless the trial function node is determined by symmetry considerations, great care must be taken when using such a method. As we have seen for the dimer HCN-He\cite{viel01,viel01b}, a non-physical node can easily lead to spurious energy levels. 
Exact calculations for $N=1$ ''dimers'' have been useful for calibrating fixed node calculations.
A more reliable approach to calculation of excited states is provided by the intrinsically more powerful and exact method of Projection Operator Imaginary Time Spectral Evolution\cite{blume97} (POITSE).
This method allows the computation of excitation energies without imposing any fixed node constraints.
The POITSE approach makes use of the inverse Laplace transform of a "projected" correlation function that leads to the spectral function 
\begin{equation}
\label{eq:kapp}
\kappa (E) = \sum_f |\langle \psi_0|\hat A|\psi_f\rangle |^2 \delta(E_0-E_f + E).
\end{equation}
Different excited energies $E_f$ are accessed by choosing the projector $\hat A$ such that the overlap $\langle \psi_0|\hat A|\psi_f\rangle$ is non-vanishing for one or a few excited states $\psi_f$.
The evaluation of the correlation decay that constitutes the transform of Eq.~(\ref{eq:kapp}) is done by making DMC side walks on the Metropolis sampling of a ground state density\cite{blume97}.
The numerical implementation used in the studies we summarize here has been described elsewhere\cite{huang01b,viel01b}.
It is based on performing the DMC side walks with pure branching, rather than the pure weights version that was used originally\cite{blume97}. 
This allows for the computation of excited states of much larger systems than studied previously.

For both fixed node and POITSE calculations, one needs to have some information about the nodal structure of the excited state of interest. 
In the case of fixed node, this is obviously extremely important since as mentioned above, the results are {\em completely} dependent on the choice of this nodal surface\cite{viel01b}.
In contrast, with POITSE, one can reach the exact energy level even with a "wrong" projector.  
This is nicely illustrated by the different behavior in $^4$He$_N$ of the three molecules studied here.
Examination of the symmetry of rotational spectra obtained in helium droplets (spherical top for SF$_6$ and linear molecule for OCS), as well as the theoretical study of rotational excitations, made for the reduced dimensionality "diatomic" SF$_6$ model\cite{blume99} indicates that the relevant excited energy levels are those in which the rotational energy is located primarily on the molecule, i.e., $J \sim j$.
One corresponding wave function for this situation in the case of a free molecule is the Wigner function $|j=1,k=0,m=0\rangle$. 
One can use this Wigner function as a trial nodal structure or, alternatively, as a projector for POITSE.
In the fixed node approximation, the results are valid only insofar as this particular nodal constraint is valid, i.e., physically correct.
In contrast, application of the POITSE method with this  Wigner projector leads to the correct energy levels no matter how weakly the true nodal structure in the cluster resembles that of the free molecule. 

For small $N$, these rotational constants derived from direct computations of the rotational energy levels of the doped cluster can be usefully compared with values obtained from rigid coupling approximations for the {\em total} helium density. 
This extreme dynamical approximation can also be described as a form of adiabatic following, although it is important to realize that the consequences of following with rigid coupling of the total helium density are different from both the adiabatic following of a hydrodynamically responding 
fluid\cite{kwon00,callegari99}, and the adiabatic following analysis for a two-fluid decomposition of the density\cite{kwon00,patel01}.

\subsection{PIMC, two-fluid, and hydrodynamic models}
The path integral Monte Carlo approach allows direct calculation of thermal averages of observables, with incorporation of the boson permutation symmetry for $^4$He and without introduction of any trial function bias.  
It is currently also the only numerical method capable of directly addressing superfluidity at non-zero temperatures.  

In PIMC one computes the average of a quantum operator over the thermal density matrix:
\begin{equation}
\avg{\hat{O}} = \frac{1}{Z} \int d\RR d\RR' \rho(\RR,\RR';\beta) \bra{\RR}\hat{O}\ket{\RR'}, 
\label{Eq_thermal}
\end{equation}
Here 
$\rho(\RR,\RR';\beta)=\bra{\RR'}e^{-\beta\hat{H}}\ket{\RR}$ is the
thermal density matrix in the position representation, $ Z =  \int d\RR \rho(\RR,\RR;\beta)$ is the partition function, and $\beta=1/k_bT$.  
The path integral Monte Carlo approach to evaluating Eq.~(\ref{Eq_thermal}) starts from the discrete Feynman path-integral expansion of $\rho(\RR,\RR';\beta)$ in terms of high temperature components $\rho(\RR_1,\RR_2;\tau)$, where $\tau=\beta / M$ constitutes an imaginary time step. 
Use of the discrete path integral expansion relies on the ability to find accurate high-temperature density matrices $\rho(\RR_1,\RR_2;\tau)$ that render evaluation of the consequent highly multi-dimensional integrals worthwhile, given the intractatibility of finding representations for the low temperature density matrix of an interacting quantum system, $\rho(\RR,\RR';\beta)$.
For $^4$He systems, the density matrix must also be symmetrized with respect to particle permutation exchanges, increasing the computational complexity.
This can be done with multi-level Metropolis sampling schemes that provides simultaneous sampling of the permutation and configuration space, allowing direct numerical study of superfluid properties at finite temperature, in addition to structural and energetic properties\cite{ceperley95}.

Path integral calculations provide estimators for thermal expectation values (Eq.~(\ref{Eq_thermal})).  
Linear response estimators for global superfluid fractions can be defined for both homogeneous and inhomogeneous systems, bulk or finite in extent\cite{ceperley95}. 
Thus quantification of a superfluid on a nanoscale is accessible from the global superfluid fraction for a pure helium droplet.
The study of molecules in helium clusters has raised the question of what the {\em local} superfluid density around a molecule is, and how this is distributed over the quantum solvation structure induced by the molecular interaction\cite{kwon00}.  
Recent work in our group has proposed a local superfluid estimator in terms of the length of the Feynman paths\cite{kwon99}.
This has been used to establish a local two-fluid model of the quantum solvation structure, based on the finding from PIMC calculations that within the first solvation shell, a local nonsuperfluid density is induced by the molecular interaction with helium\cite{kwon99,kwon00}.  
The extent of this molecular-interaction induced nonsuperfluid density depends on the strength of the interaction potential.  

This local two-fluid model has provided an atomic scale analysis of superfluid solvation\cite{kwon99}, and has also been used to analyze the rotational dynamics of the molecule\cite{kwon00}.  
The key additional concept for analysis of molecular rotation is the notion of adiabatic following of some or all of the local helium density with the molecular rotation. 
This adiabatic following, already encountered and quantified in the $T=0$ diffusion Monte Carlo calculations discussed above, leads to very different consequences for the local nonsuperfluid density and for the local superfluid density.
For the local nonsuperfluid density, adiabatic following implies an effective rigid coupling to the molecular rotation, while adiabatic following of the local superfluid density can give rise to hydrodynamic added inertia\cite{whaley98,kwon00}, analogous to the well-known added mass of macroscopic hydrodynamics.
As discussed in detail in Ref.\citelow{kwon00}, this results in a microscopic analog of the Andronikashvili experiment, with the additional distinction that the molecular rotation is quantized and does not provide a continuously varying driving force, unlike a macroscopic rotating probe.
This results in additional angular momentum constraints when the two-fluid analysis of molecular rotation is applied to interpretation of spectroscopic experiments\cite{kwon00}.  
An alternative pure hydrodynamic model is obtained as the limiting case of the two-fluid model when the entire density is assumed to be superfluid and undergoes adiabatic following\cite{callegari99,kwon00}.

\section{Applications} \label{applications}
In this section we present a summary of the various quantum Monte Carlo applications made to date for molecules embedded in $^4$He clusters.
We focus on three different molecules: SF$_6$, a slow spherical rotor, OCS, a slow linear rotor, and HCN, a fast linear rotor. 
Table \ref{Bvalues} summarizes the corresponding rotational constants in the gas phase ($B_0$) and in helium clusters ($B$), and also gives the molecular masses used in the calculations.
\begin{table}[h]
\caption[]{Experimentally rotational constants in gas phase $B_0$ and inside helium droplets $B$, in \cm. Also reported are the corresponding masses, in amu.}
\label{Bvalues}
\begin{center}
\begin{tabular}{|lllll|}
\hline
Molecule $\hspace{1cm}$& $B_0 \hspace{0.9cm}$ & $B\hspace{0.9cm}$ & Ref. &mass \\\hline
SF$_6$   & 0.091 & 0.034 &\citelow{hartmann95,harms97}& 146.0288 \\
OCS      & 0.20  & 0.073 &\citelow{grebenev00b}       & 60.07455\\
HCN      & 1.47  & 1.20  &\citelow{conjusteau00}      &27.01104 \\
\hline
\end{tabular}
\end{center}
\end{table}
\subsection{SF$_6$}
The first experimental rotational spectrum was obtained for the spherical top SF$_6$\cite{hartmann95,harms97}.
For this slow rotor, the reduction of the $B$ value inside a helium cluster is relatively large, with $B/B_0 \sim 36$\%.
This molecule was also the first for which theoretical studies of the rotational dynamics were made\cite{blume99,lee99,kwon00}.
Using the anisotropic potential energy surface of Pack and co-workers\cite{pack84}, SF$_6$ He$_N$ has been studied both with PIMC,\cite{kwon00} and with DMC in the fixed node approximation\cite{lee99}.
A POITSE study of a diatomic model of SF$_6$ has also been made\cite{blume99}.

The nonsuperfluid density of helium computed via PIMC is located in the first solvation shell and shows angular modulation structure resulting from the anisotropic interaction of helium with the octahedral SF$_6$ molecule\cite{kwon99,kwon00,toennies01}.
The energetic criterion for adiabatic following is fulfilled in this case, and application of the two-fluid dynamical model leads to a value of $B=0.033$ \cm in very good agreement with the experimental value.
Hydrodynamic calculations employing PIMC densities calculated at the experimental temperature of $T=0.3$ K show negligible added mass contribution from either the superfluid or total density\cite{kwon00}.
The average of the $Q$-factor for the dimer SF$_6$-He is found to be around 0.7\cite{patel01}.
This value is large but not equal to unity, indicating that the adiabatic following of the helium density with the molecular rotation is substantial but not complete in this system.  
The $Q$-factor is similarly less than unity in larger clusters\cite{kwon00}, implying that there is only partial adiabatic following by the density within the first solvation shell. 
This is consistent with the detailed dynamical analysis of the two-fluid model\cite{kwon00}.

\begin{figure}[h]
\begin{center}
\scalebox{0.40}{\includegraphics{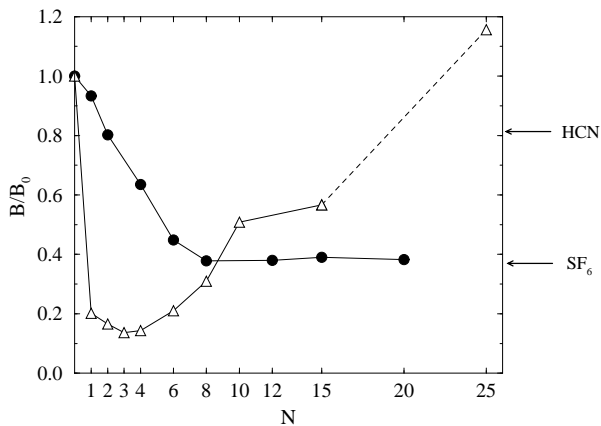}}
\end{center}
\caption[]{Evolution of $B/B_0$ with the number of helium atoms $N$ for 
SF$_6$ (circles) and HCN (triangles). The experimental values in large droplets, 36\% (SF$_6$) and 81\% (HCN), are indicated by the arrows on the right hand axis. 
For SF$_6$ the theoretical results were obtained within the fixed node approximation\cite{lee99}, and for HCN from POITSE calculations\cite{viel01b}.}
\label{Fig_B_B0}
\end{figure}

The extraction of the rotational constant $B$ has also been performed by direct computation of energy levels.
$J=1,2,3$ levels for clusters with $N=1$ to $20$ helium atoms were calculated in fixed node, employing the free molecule nodal surfaces.
For all sizes, the levels can be beautifully fit to the spherical top formula $BJ(J+1)$. 
The resulting $B$ value decreases monotonically from the gas phase value as $N$ increases from unity until it reaches a plateau at $N=8$ (Fig.~\ref{Fig_B_B0}).
An asymptotic value of $B=0.035$ \cm is obtained, in very good agreement with the experimental value extracted from spectra in large $^4$He droplets (Table \ref{Bvalues}).

In the POITSE study for SF$_6$, the octahedral SF$_6$ molecule was represented by an effective diatomic\cite{blume99}. 
The use of the original numerical implementation of POITSE employing continuous weights in the DMC sidewalks limited this study to $N=1,2$ and 3 helium atoms.
Nevertheless, the extracted $B$ values are similar to the values obtained from the full dimensionality DMC calculations within the fixed node approximation (see discussion in Ref.~\citelow{kwon00}), indicating that the free molecule nodal structure used is indeed physical in this particular case.
\subsection{OCS}
The second dopant molecule studied by quantum Monte Carlo methods is OCS, which is a linear rotor possessing a gas phase rotational constant $B_0$ twice as big as that of SF$_6$.
The increase of the moment of inertia measured in helium droplets\cite{grebenev00b} leads to a ratio $B/B_0 \sim 37$ \% that is nevertheless similar to the ratio for SF$_6$.
Because the anisotropy of the OCS-He potential\cite{higgins99} is very pronounced, we expect that adiabatic following will also be strong for this system, even though the rotor is somewhat faster than SF$_6$.
Path integral computation of the local superfluid and nonsuperfluid solvation densities show that the energetic criterion for adiabatic following is indeed fulfilled\cite{kwon00}.
Computation of the rotational constant using the two-fluid dynamical model leads to a value $B=0.067$\cm when angular momentum constraints are applied\cite{kwon00}, in good agreement with the experimental value reported in Table~\ref{Bvalues}.
For this system, the $Q$-factor computed by DMC is also around 0.7, from which we can conclude that for OCS, adiabatic following of the helium density is significant but not complete.

When a quasi-adiabatic rigid coupling analysis\cite{kwon00,quack91,paesani01} is made for a small number of helium atoms ($N<7$), this shows that as for SF$_6$, the rotational $B$ constant decreases monotonically with $N$, and that a saturation will probably occur for 5 or 6 helium atoms.  
However, as pointed out in Ref.~\citelow{kwon00}, and confirmed in Ref.~\citelow{paesani01b}, this saturation cannot be seen within the rigid coupling approximation, which continues to show a monotonic decrease of $B$ as the cluster size increases.
In contrast, the POITSE approach allows calculation of the rotational excitation without any dynamical approximation. POITSE calculations have recently allowed the saturation at small $N$ values to be demonstrated explicitly\cite{paesani01}.
A projector based on the free molecule Wigner function, $|j=1,k=0,m=0 \rangle$ produces a POITSE decay which contains only a single exponential term.
This is easily Laplace inverted using the maximum entropy approach.
The position of the single peak obtained, which we equate with $2B$, moves towards a smaller value as $N$ increases from 1 to 6.
The peak position then remains unchanged as $N$ is further increased to 10 and to 20.
The extracted saturation $B$ value of 0.07 \cm is in good agreement with the experimental value measured in much larger clusters (see Table~\ref{Bvalues}).
This saturation at a small number of helium atoms corresponding to only a fraction of the first solvation shell is similar to the situation described above for the octahedral SF$_6$ molecule\cite{lee99}.  
It similarly implies that for OCS a fraction of the first solvation shell density undergoes adiabatic following characterized by an effective rigid coupling of only that density fraction to the molecular rotation.

\subsection{HCN}
The last dopant molecule we discuss here, HCN, is also a linear rotor, but a much faster one than OCS. 
The value of $B_0$ for HCN is more than one order of magnitude larger than the values for SF$_6$ and OCS.
The reduction of rotational constant observed experimentally for HCN is also much smaller than the corresponding reduction for the two previous molecules.  
Thus, for HCN the ratio $B/B_0$ is $\sim$ 81\%\cite{conjusteau00}.
Path integral analysis concluded that both the two-fluid dynamical model and hydrodynamic models are not applicable here because of lack of efficient adiabatic following of the HCN rotation\cite{kwon00}.
This is not surprising, given the very weak anisotropy of the interaction potential for HCN with a single helium atom\cite{atkins96} and the relatively high value of the gas phase rotational constant $B_0$.

Confronted with the impossibility of applying two-fluid or hydrodynamic models in this case of a fast rotor, energy level calculations have been performed to directly access the rotational excitations in helium clusters\cite{viel01b}. 
These have been made with both fixed node and with the exact POITSE method, and the effective rotational constant $B$ extracted in each case by fitting energy level differences. 
By comparison with levels obtained from basis set calculations\cite{atkins96,viel01b}, it was found that for HCN at the very smallest cluster size ($N=1$), the nodal structure of the free molecule is not physical and leads to an erroneous energy level that does not appear in the basis set calculations.  
We have proposed that this feature derives from an unusual coincidence between the ground state energy and the potential barrier in the HCN-He system\cite{viel01b}.

Since the nodal structure of the excited state is not easily approximated for this light molecule, the POITSE approach is certainly the method of choice here.
We have shown that the spectral inversion of exponential projector correlation decays obtained from the same projector as that used for OCS (see above) leads now to spectra with two peaks\cite{viel01b}.
Excellent agreement is found while comparing the positions of these two peaks for HCN-He with results of basis set calculations, implying that this projector now overlaps with two excitations neither of which possesses the free molecule nodal structure.
This difference in the number of excitations seen in the POITSE spectra obtained from the same projector for the OCS and HCN molecules derives from the fact that these two systems lie in very different dynamical regimes, for which quite different nodal structures appear to apply.

The size dependence of the calculated $B$ value for HCN is compared with that for the heavier SF$_6$ molecule in Fig.~\ref{Fig_B_B0}. 
It is evident that molecules in the two different dynamical regimes show different saturation behavior.  
While the heavy molecule shows a monotonic decrease in $B/B_0$ to its saturation value, the light molecule first undershoots and then gradually increases to its saturation value at a significantly larger value of $N$. 
This slower saturation behavior is believed also to be related to the complex interplay between rotation and potential hindrance in the HCN-He system\cite{viel01b}. 
In both cases it is nevertheless evident that saturation to the experimental value of $B/B_0$ occurs at a cluster size $N$ that is far smaller than the experimental cluster sizes ($N \sim 4000$\cite{toennies98}), confirming that the bosonic quantum solvation effect on dopant rotations in $^4$He is primarily a local effect.

\section{Conclusion}\label{conclusion}
We have presented a comparative study of several doped helium clusters using quantum Monte Carlo techniques.
For the heavier two dopants presented here, SF$_6$ and OCS, the path integral calculations show that the quantum solvation in superfluid \4He exhibits a molecule-induced nonsuperfluid density in the first solvation shell.
The extent of this nonsuperfluid density is directly related to the strength of the impurity-helium interaction potential.
For HCN, the interaction potential is sufficiently weak that a nonsuperfluid density cannot be consistently defined.
With both finite and zero temperature techniques, we are able to calculate the extent of adiabatic following of molecular rotation by a fraction of the helium density.
The adiabatic following calculated for the two heavier molecules, SF$_6$ and OCS, is quite similar and strong, whereas for HCN the extent of adiabatic following is very small.
This strong difference between the heavy and light molecules has the consequence that neither two-fluid, pure hydrodynamic, nor quasi-adiabatic rigid coupling models apply for the computation of the rotational constant of HCN in helium clusters.

We note that rotation of the rigid body molecule is not included in the path integral studies summarized here.
While we know that the effect of molecular rotation on the helium density is limited for slow rotors, this effect is likely to be much more important for fast rotors, as was already seen with DMC densities for HCN\cite{viel01b}. 
Introduction of the molecular rotation in our path integral computations is therefore planned for future studies.
Moreover, the weak coupling angular momentum constraint $J\simeq j$ appears to be invalid for HCN, at least for the small cluster sizes, as demonstrated by the marked inaccuracy of fixed node calculations employing free molecule nodal surfaces.
All these points, plus the fact that spectra obtained by the exact POITSE method using the same projector are nevertheless different for the two linear molecules HCN and OCS, underline the marked differences between the rotational dynamics of the fast linear rotor HCN from that of the heavier molecules SF$_6$ and OCS in helium clusters.

\section*{Acknowledgments}
We acknowledge financial support from the National Science Foundation through NSF grant CHE-9616615.
Supercomputer time was made available through a National Partnership for Advanced Computational Infrastructure (NPACI) program of NSF, administered by the San Diego Supercomputer Center.  
We would like to thank Patrick Huang for stimulating discussions and Francesco Paesani for permission to discuss the OCS POITSE results prior to publication.


\begin{thebibliography}{10}

\bibitem{toennies98}
J.~P. Toennies and A.~F. Vilesov, Ann. Rev. Phys. Chem. {\bf 49},  1  (1998).

\bibitem{toennies01}
J.~P. Toennies, A.~F. Vilesov, and K.~B. Whaley, Physics Today {\bf 54},  31
  (2001).

\bibitem{kwon00}
Y. Kwon {\it et~al.}, J. Chem. Phys. {\bf 113},  6469  (2000).

\bibitem{conjusteau00}
A. Conjusteau {\it et~al.}, J. Chem. Phys. {\bf 113},  4840  (2000).

\bibitem{whaley98}
K.~B. Whaley, {\em Advances in Molecular Vibrations and Collision Dynamics,
  vol. III} (ed. J. Bowman and Z. Ba\u{c}i\'c, Academic Press, JAI Press Inc., 1998),
  pp.\ 397--451.

\bibitem{viel01}
A. Viel, M.~V. Patel, P. Niyaz, and K.~B. Whaley, Comp. Phys. Com.  in press
  (2001).

\bibitem{lee99}
E. Lee, D. Farrelly, and K.~B. Whaley, Phys. Rev. Lett. {\bf 83},  3812
  (1999).

\bibitem{patel01}
M.~V. Patel, A. Viel, F. Paesani, and K.~B. Whaley, to be published  .

\bibitem{viel01b}
A. Viel and K.~B. Whaley, J. Chem. Phys. {\bf 115},  XX  (2001).

\bibitem{blume97}
D. Blume, M. Lewerenz, P. Niyaz, and K.~B. Whaley, Phys. Rev. E {\bf 55},  3664
   (1997).

\bibitem{huang01b}
P. Huang, A. Viel, and K.~B. Whaley,  in {\em Recent advances in Quantum Monte
  Carlo Methods, Part II} (WORLD Scientific, New York, 2001).

\bibitem{blume99}
D. Blume, M. Mladenovi\'c, M. Lewerenz, and K.~B. Whaley, J. Chem. Phys. {\bf
  110},  5789  (1999).

\bibitem{callegari99}
C. Callegari {\it et~al.}, Phys. Rev. Lett. {\bf 83},  5058  (1999).

\bibitem{ceperley95}
D.~M. Ceperley, Rev. Mod. Phys. {\bf 67},  279  (1995).

\bibitem{kwon99}
Y.~K. Kwon and K.~B. Whaley, Phys. Rev. Lett. {\bf 83},  4108  (1999).

\bibitem{hartmann95}
M. Hartmann, R.~E. Miller, A.~F. Vilesov, and J.~P. Toennies, Phys. Rev. Lett.
  {\bf 75},  1566  (1995).

\bibitem{harms97}
J. Harms {\it et~al.}, J. Mol. Spectrosc. {\bf 185},  204  (1997).

\bibitem{grebenev00b}
S. Grebenev {\it et~al.}, J. Chem. Phys. {\bf 112},  4485  (2000).

\bibitem{pack84}
R.~T. Pack, E. Piper, G.~A. Pfeffer, and J.~P. Toennies, J. Chem. Phys. {\bf
  80},  4940  (1984).

\bibitem{higgins99}
K. Higgins and W.~H. Klemperer, J. Chem. Phys. {\bf 110},  1383  (1999).

\bibitem{quack91}
M. Quack and M.~A. Suhm, J. Chem. Phys. {\bf 95},  28  (1991).

\bibitem{paesani01}
F. Paesani, F.~A. Gianturco, A. Viel, and K.~B. Whaley, to be published  .

\bibitem{paesani01b}
F. Paesani, F.~A. Gianturco, and K.~B. Whaley, Europhys. Lett.  in press
  (2001).

\bibitem{atkins96}
K.~M. Atkins and J.~M. Hutson, J. Chem. Phys. {\bf 105},  440  (1996).

\end{thebibliography}
\end{document}